# Picosecond opto-acoustic interferometry and polarimetry in high-index GaAs

A. V. Scherbakov,[1*] M. Bombeck,[2] J. V. Jäger,[2]  A. S. Salasyuk,[1] T. L. Linnik,[3]
V. E. Gusev,[4] D. R. Yakovlev,[1,2]  A. V. Akimov,[1,5] and M. Bayer[1,2]

[1]*Ioffe Physical-Technical Institute, Russian Academy of Sciences, 194021 St. Petersburg, Russia*
[2]*Experimentelle Physik 2, Technische Universität Dortmund, D-44227 Dortmund, Germany*
[3]*Department of Theoretical Physics, V.E. Lashkaryov Institute of Semiconductor Physics, 03028 Kyiv, Ukraine*
[4]*IMMM, UMR-CNRS 6283, LUNAM, Université du Maine, Av. O. Messiaen, 72085 Le Mans, France*
[5]*School of Physics and Astronomy, University of Nottingham, Nottingham NG7 2RD, UK*

## ABSTRACT

By means of a metal opto-acoustic transducer we generate quasi-longitudinal and quasi-transverse picosecond strain pulses in a (311)-GaAs substrate and monitor their propagation by picosecond acoustic interferometry. By probing at the sample side opposite to the transducer the signals related to the compressive and shear strain pulses can be separated in time. In addition to conventional monitoring of the reflected probe light intensity we monitor also the polarization rotation of the optical probe beam. This polarimetric technique results in improved sensitivity of detection and provides comprehensive information about the elasto-optical anisotropy. The experimental observations are in a good agreement with a theoretical analysis.

## 1. Introduction

In 1984 *Thomsen et al.* [1,2] developed the method of generation and detection of ultrashort coherent acoustic pulses in solids by femtosecond laser excitation. Thereby the energy of a short optical pulse being absorbed in the near-surface layer of a solid medium is converted into coherent lattice vibrations. The resulting elastic excitation is a strain pulse of picosecond duration with amplitude up to $10^{-3}$ and $\sim 10 \div 100$-nm spatial extension, propagating through the crystal with sound velocity [2]. The excitation spot diameter is typically much larger than the penetration depth of light in the medium and thus the propagating pulse may be considered as a superposition of plane acoustic waves with wave vectors normal to the excited surface. This coherent wavepacket has a broad acoustic spectrum with frequencies up to hundreds of gigahertz (GHz). The composition by such high frequencies represents the main advantage of the femtosecond optical excitation compared to frequency-limited conventional techniques using piezoelectric transducers.

The detection of picosecond strain pulses is based on time-resolved monitoring of the strain-modulated reflectivity [2]. In an opaque material this detection occurs in the near-surface region providing information about the strain pulse amplitude and temporal shape. Shortly after development this method was extended by the technique of picosecond acoustic interferometry, which is suitable also for transparent media [3-5]. Here the intensity of coherent light reflected at an acoustic wavepacket propagating in the optically transparent medium oscillates in time with the frequency given by

$$f_B = \frac{2 \upsilon n}{\lambda} \cos \theta, \qquad (1)$$

where $\upsilon$ is the sound velocity, $n$ is the refractive index of the medium, $\lambda$ is the wavelength of the reflected light in vacuum and $\theta$ is the incident angle of light inside the medium. The origin of the oscillations is dynamical interference of light partly reflected at the crystal surface with light reflected at the propagating strain pulse. These oscillations are commonly called Brillouin oscillations because $f_B$ is equal to the frequency shift in corresponding Brillouin scattering spectra for specular reflected light. Monitoring the Brillouin oscillations with picosecond time resolution allows one to obtain comprehensive information about the elastic, optical, electronic and mechanical properties of solid-state materials and structures [6-15], liquids [16-18] and biological objects [19,20].

In both cases, when the medium is either strongly absorbing or transparent for the probe light, its interaction with a coherent acoustic wavepacket is result of the elasto-optical effect, i.e. the strain-induced changes of the permittivity. The elasto-optical effect in general causes a medium to be optically anisotropic [21]. Thus, the changes of the intensity of the reflected light induced by the picosecond strain pulse become dependent on the polarization of the probe beam. Only for high-symmetry conditions, representing quite some restrictions, the strain-induced modulation of reflectivity remains insensitive to the probe light polarization: this is the case, for instance, for normally incident light scattered on a longitudinal acoustic wavepacket propagating along a high-symmetry crystallographic direction. Any reduction of symmetry, however, will make the optical probing polarization-sensitive.

The anisotropy of the elasto-optical interaction becomes especially important in experiments with picosecond shear strain pulses. The reflectivity signal induced by a transverse acoustic wavepacket depends on the polarization of the probe light. This effect has been addressed theoretically [21-25] and examined experimentally [23-26]. Experiments on near-surface detection of shear strain pulses have demonstrated also the possibility to extend the picosecond acoustic techniques by the methods of transient optical polarimetry, i.e. time-resolved monitoring of the ellipticity of the reflected, originally linearly-polarized probe light or the rotation of its polarization plane [27]. This possibility has been also theoretically analyzed [28].

Generation of shear strain pulses with picosecond duration, which has remained a challenge so far, may significantly extend the applications of high-frequency acoustics due to the smaller velocities and wavelengths of transverse acoustic waves. In related experiments [23-27] shear and compressive acoustic modes were generated simultaneously. They have different sound velocities, $v$, resulting in different frequencies in the Brillouin signal, see Eq. (1) [23,25,26]. As generation and probing were performed in these experiments at the same spot of the studied sample one could not separate the contribution of compressive and shear acoustic modes in time, resulting in complex temporal signals. Thus, the shear component remained weak on the dominant background of the compressive contribution and therefore arduous for detection. This severely limits applications of picosecond acoustic interferometry using shear acoustic pulses. Moreover, in all picosecond acoustic interferometry experiments, where Brillouin oscillations were measured, only intensity changes were monitored, while polarimetric techniques were not implemented so far.

The goal of the present work is to extend picosecond acoustic interferometry by polarimetric techniques and distinguish clearly the optical probe signals resulting from compressive and shear strain pulses by using a remote geometry. In this geometry generation and probing are clearly separated in space. In our experiments we generate quasi-longitudinal (QLA) and quasi-transverse (QTA) strain pulses in a high-index GaAs substrate using a metal optoelastic transducer. The semiconductor substrate and the associated optical probe wavelength allow us to attain a large penetration depth for the probe light, thus, opening the possibility to detect the propagating acoustic wavepacket far away from the sample surface. Thus, we consider a material transparent for the probe light and focus only on the picosecond interferomic signal possessing Brilluoin oscillations with frequency given by Eq. (1). By comparing probe signals measured with traditional picosecond acoustic interferometry to signals obtained by polarimetric we show that the latter technique shows a much higher sensitivity to picosecond shear strain pulses and strongly depends on the orientation of the probe light polarization relative to the crystallographic axes of the sample.

## 2. Experiment

Figure 1(a) shows the experimental scheme in the chosen coordinate frame with the *x*, *y* and *z* axes along the $[\bar{2}33]$, $[0\bar{1}1]$ and $[311]$ crystallographic directions, respectively. The pump pulse taken from an amplified Ti:Sapphire laser (800-nm wavelength, 200 fs duration of the pulses) excites an aluminum film with 100-nm thickness deposited on the polished backside of a (311)-GaAs substrate of 100-μm thickness. The laser beam is focused to a round spot of 100-μm diameter (full width at half maximum of the Gaussian-shaped radial power distribution) providing an excitation energy density of $W=4$ mJ/cm$^2$. The strain pulse generation in the acoustically isotropic metal film by femtosecond optical excitation is well studied in literature [2,29-32]. The ultrafast thermal expansion of the film leads to injection of strain pulse into the GaAs substrate, where it propagates further normal to the surface towards the opposite side.

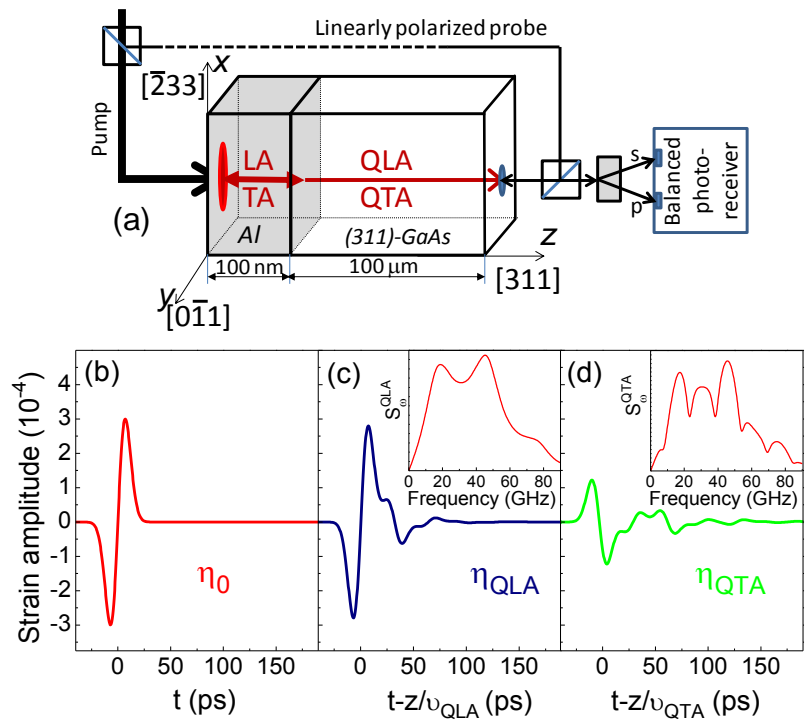

Fig. 1. (a) Experimental schematic; (b) temporal profile $\eta_0(t)$ of the initial LA strain pulse generated in the Al film; (c) and (d) temporal profiles $\eta_p(t,z)$ of the QLA (c) and QTA (d) strain pulses injected into the (311)-GaAs substrate. The insets in (c) and (d) show the respective spectral amplitudes of the injected strain pulses.

The experiments are performed in a helium cryostat at 10 K temperature. This allows the high-frequency (up to 1 THz) components of the acoustic wavepacket to reach the surface opposite to the Al transducer after propagation through the 100-μm GaAs substrate [33]. There are three acoustic modes, which propagate along the [311] crystallographic direction in GaAs: the QLA, QTA and pure transverse (TA) modes with calculated sound velocities, $v_p$ ($p$=QLA, QTA or TA), of 5.1, 2.9 and 3.2 km/s, respectively [34]. The strain pulses injected from the Al transducer into the GaAs substrate along this direction are combinations of these modes, but due to the different sound velocities of the modes they segregate in time into isolated pulses of particular polarization. The propagation times through the substrate $t_p = d/v_p$, where $d$=100 μm is the GaAs substrate thickness, are 19.6 and 34.5 ns for the QLA and QTA modes, respectively. Pure TA modes are not generated due to the symmetry selection rules.

By optical probing at the side opposite to the aluminum transducer we study the interaction of light with an isolated strain pulse of certain polarization in time. The probe pulse is split from the same laser source as the pump pulse. The probe beam is linearly polarized, hits the sample normal to the surface and is focused to a spot of 50-μm diameter opposite to the pump spot. The delay line for the probe beam is adjusted such that the time delay between the pump and probe pulses is equal to the corresponding propagation times $t_p$. The variable delay of the pump beam provides sub-picosecond time resolution for monitoring the temporal evolution of the reflectivity of the probe pulse. In the experiments we monitor the modulation of the reflected probe pulse intensity $\Delta I(t)/I_0$ ($I_0$ is the reflected probe intensity in absence of the strain pulses) and the strain-induced rotation of the probe pulse polarization plane $\Delta\psi(t)$. For monitoring the intensity changes $\Delta I(t)/I_0$ we use a single photodiode. In the measurements of $\Delta\psi(t)$ a balanced detection scheme in used: the reflected probe beam is split into two orthogonally $s$- and $p$-polarized components balanced to equal intensities in absence of the strain pulse. Both signals $\Delta I(t)/I_0$ and $\Delta\psi(t)$ are measured at several angles $\psi_0$=0, $\pi/4$, $\pi/2$ and $3\pi/4$ between the equilibrium orientation of the probe polarization plane and the $x$-axis.

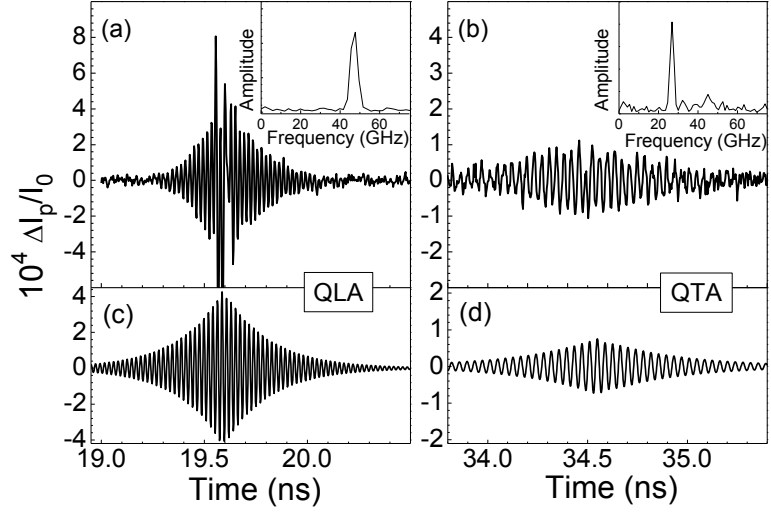

Figure 2 (a) and (b) Experimental signals $\Delta I_p(t)/I_0$ measured around the time delays corresponding to the propagation time of the QLA and QTA pulses, respectively. The insets show the FFT spectra of the measured signals obtained in time windows of 0.5 ns duration starting at $t_p$+0.1 ns. (c) and (d) Calculated signals $\Delta I_p(t)/I_0$ induced by the QLA (c) and QTA (d) strain pulses.

Panels (a) and (b) in Fig. 2 show the signals $\Delta I_p(t)/I_0$ ($p$ indicates the strain pulse mode, QLA or QTA) measured at the corresponding QLA or QTA delays. No signal is observed at the delay corresponding to the propagation time of the TA pulse, indicating that this mode is not excited as expected. Both signals show GHz oscillations, which are modulated by an almost symmetric envelope function. The center of this function at $t=t_p$ indicates the time, when the strain pulse is reaching the open surface of the GaAs substrate and becomes reflected there. The total signal duration (~1 ns) is governed by the light absorption and is determined by the time required for the strain pulse to travel the double (forward and backward) penetration depth (~2 μm) of the probe light. Obviously, the longer signal is induced by the QTA pulse [compare panels (a) and (b) of Fig. 2] due to its lower sound velocity. The oscillations detected in $\Delta I_p(t)$ have frequencies determined by Eq. (1) and actually are the Brillouin oscillations governed by the picosecond acoustic interferometry. For $n=3.6$ and $\lambda=800$ nm at normal incidence ($\theta=0$) as well as the corresponding values of $\upsilon_p$ the frequencies calculated after Eq. (1) are $f_B^{QLA}=46$ GHz and $f_B^{QTA}=27$ GHz for the QLA and QTA modes, respectively. These values are in perfect agreement with the fast Fourier transform (FFT) spectra of the signals, shown in the insets of Fig. 2 [35]. Comparing the panels (a) and (b) of Fig. 2 one sees that the amplitude of $\Delta I_{QLA}(t)$ for the QLA mode is about five times larger than the amplitude of $\Delta I_{QTA}(t)$ for QTA. The signals $\Delta I_{QLA}(t)$ and $\Delta I_{QTA}(t)$ are independent on $\psi_0$.

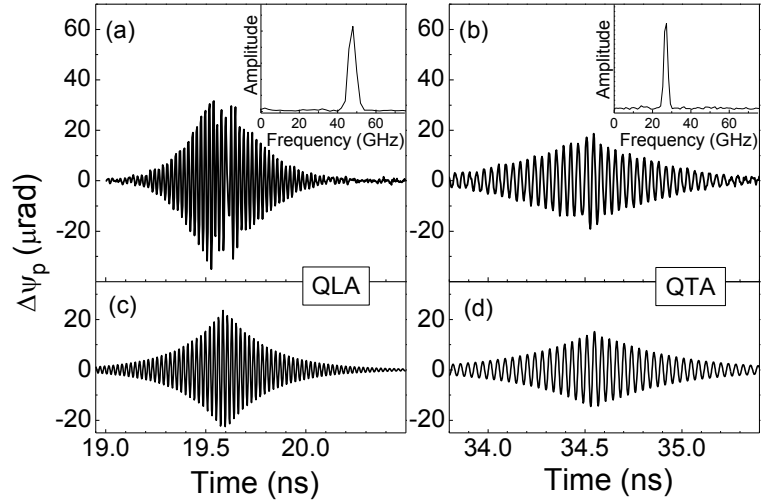

Figure 3 (a) and (b) Experimental signals $\Delta\psi_p(t)$ measured around time delays corresponding to the propagation time of the QLA (a) and QTA (b) pulses through the substrate, respectively. The insets show the FFT spectra of the measured signals obtained in time windows of 0.5-ns duration starting at $t_p+0.1$ ns. (c) and (d) Calculated signals $\Delta\psi_p(t)$ induced by the QLA (c) and QTA (d) strain pulses.

Figures 3a and 3b show the modulations of the probe light polarization induced by the strain pulses, $\Delta\psi_p(t)$, measured at the same delays as in Fig. 2. The signal to noise ratio is improved by balancing the laser intensity noise in the detection scheme. Thus, the polarimetric signal induced by the QTA pulse is much more pronounced in comparison with the intensity signals $\Delta I_{QLA}(t)$ and $\Delta I_{QTA}(t)$. Both $\Delta\psi_{QLA}(t)$ and $\Delta\psi_{QTA}(t)$ demonstrate behaviors similar to those observed for the intensity modulation, but the difference in their amplitudes is smaller. Contrary to $\Delta I_p(t)$, the polarimetric signals $\Delta\psi_p(t)$ demonstrate strong dependences on $\psi_0$ for the strain pulses of both polarizations. This is demonstrated in Fig. 4, which shows the signals $\Delta\psi_{QLA}(t)$ and $\Delta\psi_{QTA}(t)$ measured at different $\psi_0$. One can see that when the polarization plane of the probe light is parallel to the [2̄33] or [01̄1] crystallographic directions the signals are extremely weak.

The Brillouin oscillations detected in $\Delta\psi_p(t)$ and the strong dependence of the polarimetric signal on the orientation of the probe polarization plane, indicating strain-induced optical anisotropy, are the main experimental observations of this work. Below we present a theoretical analysis, which takes into account strain pulse generation and detection of the Brillouin oscillations in a low-symmetry GaAs substrate.

## 3. Theoretical analysis and discussion

The generation of the strain pulse takes place in the Al film deposited on the GaAs substrate. The thickness of the film (100 nm) is much larger than the penetration depth of the pump light (~10 nm), and assuming subsonic thermal diffusion we may consider two processes independently: generation of the initial strain pulse in the Al film and then injection of this pulse into GaAs [25,31].

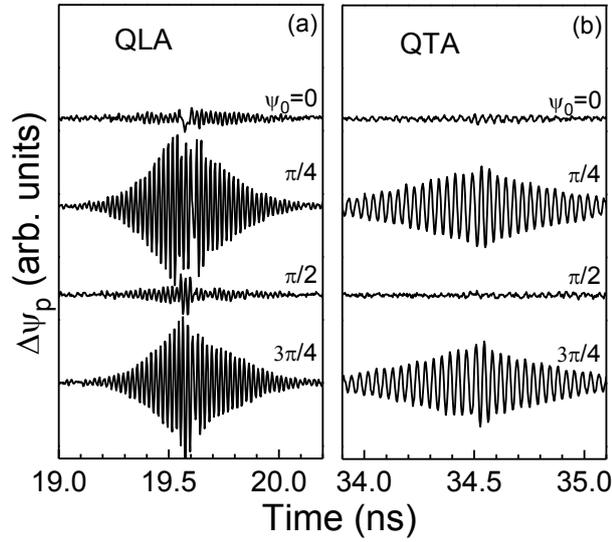

Figure 4. Signals $\Delta\psi_p(t)$ measured for four different orientations of the probe beam polarization plane around $t_{QLA}$ (a) and $t_{QTA}$ (b) pulses, respectively.

The hot carriers, optically generated in the narrow near surface region of aluminum (~10 nm), diffuse into the film and pass their energy to the lattice. The combined effect of the increased lattice temperature and the nonequilibrium carriers initiates the mechanical stress generating the strain pulse [2,29-32]. The Al film is polycrystalline and behaves as an elastically isotropic medium, thus the initially generated strain pulse propagating normally to the surface is purely longitudinal (LA). From previous studies it is known, that such a pulse is bipolar and with good accuracy can be approximated by the derivative of a Gaussian function [36]. Figure 1b shows the calculated initial strain pulse $\eta_0(t-z/\upsilon_{LA})$, where $\upsilon_{LA}$=6.3 km/s is the longitudinal sound velocity in Al [37].

The initial step of strain pulse generation in the Al film is followed by transmission of the "seed" strain pulse through the interface between the Al film and the acoustically anisotropic (311) GaAs substrate. In the coordinate frame shown in Fig. 1a the QLA, QTA and TA modes, propagating along [311] (i.e. the $z$-direction), have the unit displacement vectors $\mathbf{e}^{QLA}=(u_0,0,\sqrt{1-u_0^2})$, $\mathbf{e}^{QTA}=(\sqrt{1-u_0^2},0,-u_0)$ and $\mathbf{e}^{TA}=(0,1,0)$, where $u_0$=0.165 [34]. The mixed character of the QLA and QTA waves gives rise to their coupling with the pure longitudinal wave generated in the Al film, while the pure TA mode is not excited. The related strain components of the QLA and QTA pulses may be written as:

$$u_{zz}^p(t,z) = e_z^p \eta_p(t-z/\upsilon_p), \quad u_{xz}^p(t,z) = \frac{1}{2}e_x^p \eta_p(t-z/\upsilon_p), \quad (p=\text{QLA,QTA}). \quad (2)$$

The functions $\eta_p(t-z/\upsilon_p)$ describe the spatio-temporal evolution of the strain pulses and are determined by the "seed" pulse profile and its transmission through the Al/GaAs interface. The acoustic mismatch between aluminum and GaAs leads to partial reflection of the initial pulse at the interface. This reflection together with the acoustic mode conversion leads to appearance of a transverse pulse in the Al film ($\upsilon_{TA}$=3.1 km/s in Al [37]) that is polarized along the x-axis in addition to the pure LA pulse. These two pulses propagate in the Al film back towards the surface with their individual sound velocities and after reflection there back towards GaAs. Then the transmission/reflection process accompanied by mode conversion repeats again. This complicated process of multiple injections may be addressed quantitatively in the frequency domain using standard elasticity theory, as described in Appendix A.

Figure 1(c) and 1(d) show the calculated profiles $\eta_p(t-z/\upsilon_p)$ of the QLA and QTA strain pulses injected into the GaAs substrate. The QLA pulse is almost pure longitudinal: $u_{zz}^{QLA}/u_{xz}^{QLA}=11$ and the amplitude of its longitudinal component achieves the value of $2.8 \times 10^{-4}$ for $W$=4 mJ/cm$^2$ (see Appendix A). In the studied structure the QLA profile is determined by the shape of the LA pulse initially generated in the Al film and its multiple reflections with negligible influence of the TA pulse in the aluminum film. The QTA pulse possesses more of a mixed character with a dominating shear component ($u_{xz}^{QTA}/u_{zz}^{QTA}=2.9$) of rather small amplitude: the peak value of $u_{xz}^{QTA}$ is only $0.6 \times 10^{-4}$. The QTA temporal profile is more complicated, because it is affected by multiple reflections of both the LA and TA pulses. The insets of Figs. 1 (c) and 1 (d) show the respective spectral amplitudes $S_\omega^p$, which consists of several peaks. The spectral positions of the peaks are determined by the interference conditions for the acoustic waves in the 100-nm aluminum film: only LA waves are relevant in the spectrum of the QLA pulse, while both LA and TA waves contribute in the spectrum of the QTA pulse.

The QLA and QTA strain pulses injected into the substrate perturb the dielectric permittivity of GaAs due to the photo-elastic effect [38]. The strain modifies the permittivity according to $\delta\varepsilon_{ij}(t,z)=\varepsilon_0^2 p_{ijkl}u_{kl}(t,z)$ where $\varepsilon_0$ is the unperturbed dielectric permittivity, which is isotropic in GaAs, and $p_{ijkl}$ are the components of the photo-elastic tensor, written in the [311]-coordinate system. Since in our case there are only two nonzero strain components $u_{xz}(t,z)$ and $u_{zz}(t,z)$ for the geometry used in the experiments only two components, $\delta\varepsilon_{xx}(t,z)$ and $\delta\varepsilon_{yy}(t,z)$, have nonzero values. In the low-symmetry case these deviations are not equal, and the medium becomes optically anisotropic. Considering the spatio-temporal evolution of the strain we have to take into account that the strain pulses are reflected with $\pi$-phase shift at the free surface of the GaAs substrate. Then using the Maxwell equations in linear to strain approximation we come to the expressions for the relative strain-induced perturbation of the probe pulse intensity and polarization plane rotation (see Appendix B):

$$\frac{\Delta I(t)}{I_0} = \mathrm{Re}\left\{\frac{ik_0^2 T_{EM}}{R_{EM}\,(k_0 + k\ )}\int\limits_{-\infty}^{0} dz e^{-2ikz}\Big[\delta\varepsilon_{xx}(t,z) + \delta\varepsilon_{yy}(t,z) + [\delta\varepsilon_{xx}(t,z) - \delta\varepsilon_{yy}(t,z)]\cos 2\psi_0\Big]\right\}, \quad (3)$$

$$\Delta\psi(t) = \mathrm{Re}\left\{\frac{ik_0^2 T_{EM}\sin 2\psi_0}{2R_{EM}\,(k_0 + k\ )}\int\limits_{-\infty}^{0} dz \cdot e^{-2ikz}[\delta\varepsilon_{yy}(t,z) - \delta\varepsilon_{xx}(t,z)]\right\}. \quad (4)$$

Here $k=nk_0+i\gamma$ and $k_0=2\pi/\lambda$ are the complex wave number of light in the substrate with refractive index $n$ and in vacuum, respectively, $2\gamma$ is the absorption coefficient, $R_{EM}$ and $T_{EM}$ are the complex reflection and transmission coefficients equal to $T_{EM}=2k_0(k+k_0)^{-1}$ and $R_{EM}=(k_0-k)(k+k_0)^{-1}$. In the numerical calculations we use the value $\gamma=2.2\times10^6$ m$^{-1}$ [39].

The values of the photoelastic constants for GaAs are known [40] and allow us to get direct expressions for the dielectric permittivity perturbations:

$$\delta\varepsilon_{xx}(t,z) + \delta\varepsilon_{yy}(t,z) = -7.6u_{xz}(t,z) + 29u_{zz}(t,z),$$
$$\delta\varepsilon_{xx}(t,z) - \delta\varepsilon_{yy}(t,z) = 2.9u_{xz}(t,z) - 1.8u_{zz}(t,z). \quad (5)$$

Equations (5) show that both the $u_{xz}$ and $u_{zz}$ components of the strain pulses contribute to the detected signals. Using the Eqs. (2-5) we come to solutions of the form:

$$\Delta I_{QLA}(t)/I_0 = [28 - 1.5\cos(2\psi_0)]Q_{QLA}(t), \qquad \Delta I_{QTA}(t)/I_0 = [-8.5 + 1.7\cos(2\psi_0)]Q_{QTA}(t),$$
$$\Delta\psi_{QLA}(t) = 0.75\sin(2\psi_0)Q_{QLA}(t), \qquad \Delta\psi_{QTA}(t) = -0.85\sin(2\psi_0)Q_{QTA}(t), \quad (6)$$

with

$$Q_p(t) = \frac{2k_0}{(n^2-1)}\upsilon_p S_\omega^p \cos[2\pi f_B^p(t_p-t) + \varphi_\omega^p]\exp(-2\gamma\upsilon_p\,|\,t-t_p\,|). \quad (7)$$

Here $S_\omega^p$ and $\varphi_\omega^p$ are the absolute value and phase of the spectral amplitude, which is complex in the most general case (see Appendix A), at the Brillouin frequency determined by Eq. (1). Assuming $\varphi_\omega^p = 0$, the Brillouin signals achieve maximal amplitude at the time $t=t_p$, when the center of the strain pulse reaches the open surface [41]. The oscillations decay exponentially in time (for both positive and negative time delays) with time constant $\gamma\upsilon_p/2$ determined by the penetration depth of the probe light and the sound velocity.

Panels (c) and (d) of Figure 2 show the calculated temporal evolutions of $\Delta I_p(t)/I_0$ induced by the QLA and QTA strain pulses. We see good quantitative agreement between calculated and measured signals. The experimental signal $\Delta I_{QLA}(t)/I_0$ has several peculiarities (narrow peaks and deeps) around $t_{QLA}=19.6$ ns, which are not seen in the calculated signal. These peculiarities may be due to nonlinear effects, e.g. formation of a shock wave or picosecond acoustic solitons [42]. These effects may accompany the propagation of the compressive strain pulse through the substrate and modify its shape resulting in complex signal $\Delta I_{QLA}(t)/I_0$ [41], but are not taken into account in the calculations. We also see a good agreement between the measured [Figs. 3(a) and 3(b)] and calculated [Figs. 3(c) and 3(d)] polarimetric signals.

The pre-factors in Eq. (6) determine the angular dependences of the detected signals. The angular dependence of $\Delta\psi_p(t)$ shows: no rotation when the polarization plane of the incident probe pulse is parallel to the [$\bar{2}33$] or [$0\bar{1}1$] direction ($\psi_0=\pi l/2$), and the maximal signal should be detected at $\psi_0=(2l+1)\pi/4$ ($l=0,1,2\ldots$). This behavior is in full agreement with our experimental observations: at $\psi_0=0$ and $\pi/2$ $\Delta\psi_p(t)$ is very weak.

The dependence of $\Delta I_p(t)/I_0$ on $\psi_0$ originates from the anisotropic polarization-depended term in Eq. (6) and is much weaker than for $\Delta\psi_p(t)$. It is expected that at $\psi_0=\pi/2$ or $3\pi/2$ we should detect the maximal intensity modulation, while at $\psi_0=0$ or $\pi$ the detected signals is minimal. However, the maximal relative amplitude changes of $\Delta I_p(t)/I_0$ measured at different angles are 5% for the QLA pulse and 20% for the QTA pulse. Expressed in absolute values, they are less than $2\times10^{-5}I_0$ and arduous for detection [see the signals and their amplitudes in Figs. 2 (a) and 2(b)]. Thus, the angular dependence of $\Delta I_p(t)$ is not resolved in the experiment.

## 4. Conclusions

To summarize our experimental observations and theoretical analysis the optical anisotropy induced by the strain pulse propagating in GaAs along a high-index direction results in Brillouin oscillations of the polarization plane rotation of the probe light reflected at the propagating pulse. In the particular low-symmetry case examined in our work this effect is strong and results in a higher sensitivity for polarimetric probing than in conventional picosecond acoustic interferometry where the modulation of the probe intensity is monitored. We operate with strain pulses of mixed polarization containing both compressive and shear strain components which are well separated in time due to the remote geometry in the experiments.

We want to point out also our analysis of the strain pulse generation. We have calculated the spatio-temporal profiles of the QLA and QTA strain pulses for a generation scheme with an optoelastic transducer on the high-index semiconductor substrate and estimated the strain pulse amplitudes and durations. The reasonability of our analysis is confirmed by the perfect quantitative agreement between calculated and measured signals, which are determined by the strain pulses parameters. The proposed scheme allows utilizing shear strain pulses for ultrafast manipulation of various excitations in nanostructures and thin films, since they may be easily integrated with a semiconductor substrate. An example is the recent magneto-acoustic experiments with shear strain pulses in ferromagnetic (311)-(Ga,Mn)As layer [43].

**Acknowledgements**

The work was supported by the Deutsche Forschungsgemeinschaft (BA 1549/14-1), the Russian Foundation for Basic Research (11-02-00802), the Russian Academy of Science, the State Fund for Fundamental Researches in the frame of the program SFFR-DFG (39.2/002).

## Appendix A. Generation and injection of the strain pulses

The hot carriers, optically generated in the narrow near surface region of the Al film, diffuse into the film and pass their energy to the lattice. The combined effect of the increased lattice temperature and nonequilibrium carriers initiates the mechanical stress generating the strain pulse. The Al film structure is polycrystalline, and we can treat it as an elastically isotropic medium. In this case the generated strain pulse is pure LA. In our case, however, the TA component appears in the film as well as a result of mode conversion under reflection at the interface with the anisotropic (311) GaAs substrate. To calculate the QLA and QTA strain pulses injected in the (311)-GaAs substrate we start from the standard elasticity equations [44] for the temporal Fourier components of the displacement **u** in the aluminum film. In our case only the components $u_z(\omega,z)$ and $u_x(\omega,z)$ are nonzero and they are determined by the equations:

$$-\rho_{Al}\omega^2 u_j(\omega,z) = \frac{d\sigma_{jz}}{dz},$$

$$\sigma_{zz} = \upsilon_{LA}^2 \rho_{Al}\frac{du_z(\omega,z)}{dz} + G(\omega,z), \quad \sigma_{xz} = \upsilon_{TA}^2 \rho_{Al}\frac{du_x(\omega,z)}{dz}, \tag{1a}$$

where $G(\omega,z)$ is the Fourier component of the laser pulse induced stress in the Al film, $\sigma_{ij}$ are the components of stress tensor, $\upsilon_{LA,TA}$ and $\rho_{Al}$ are the longitudinal and transverse sound velocities and the density of Al, respectively.

After the transmission of the initial LA strain pulse through the interface, two strain pulses, QLA and QTA, propagate further in the substrate and are characterized by the complex spectral amplitudes $\tilde{S}_\omega^{QLA}$ and $\tilde{S}_\omega^{QTA}$, respectively:

$$u_{xz}^p(\omega,z) = \frac{i}{2}e_x^p \tilde{S}_\omega^p \exp(iq_p(z-d_0)), \qquad u_{zz}^p(\omega,z) = ie_z^p \tilde{S}_\omega^p \exp(iq_p(z-d_0)). \tag{2a}$$

Here $q_p = \omega/\upsilon_p$ ($p$=QLA and QTA) are the wave numbers and $d_0$ is the thickness of the Al film. The solutions for $\tilde{S}_\omega^{QLA}$ and $\tilde{S}_\omega^{QTA}$ can be found by solving Eq.(1a) for $0 < z < d_0$ and imposing the standard mechanical boundary conditions at the interface and at the free surface of the film assuming that the function $G(\omega,z)$ decays away from the film surface and is equal to zero at the interface. It is convenient to express the results via the reflection $R_{jl}$ and transmission $T_{jl}$ coefficients of both the TA and LA strain pulses at the Al/(311) GaAs interface. The $R_{jl}$ ($T_{jl}$) coefficients give the relations between the strain amplitude of the reflected (transmitted) mode $l$ and the amplitude of the mode $j$ incident on the interface. For the parameters of the Al film and GaAs [37,39] we use: $T_{LA,QLA}\approx0.931$, $T_{LA,QLA}\approx0.051$, $T_{LA,QTA}\approx-0.407$, $T_{TA,QTA}\approx0.730$, and $R_{LA,LA}\approx0.222$, $R_{TA,LA}\approx0.014$, $R_{LA,TA}\approx0.124$, $R_{TA,TA}\approx0.306$. The solution for the spectral amplitudes has the form:

$$\begin{pmatrix}\tilde{S}_\omega^{QLA}\\\tilde{S}_\omega^{QTA}\end{pmatrix}=S_0(\omega)\frac{e^{iq_{LA}d_0}}{\Delta}\left[\begin{pmatrix}T_{LA,QLA}\\T_{LA,QTA}\end{pmatrix}(1+R_{TA,TA}e^{2iq_{TA}d_0})-\begin{pmatrix}T_{TA,QLA}\\T_{TA,QTA}\end{pmatrix}R_{LA,TA}e^{2iq_{TA}d_0}\right],$$

$$\Delta=(1+R_{LA,LA}e^{i2q_{LA}d_0})(1+R_{TA,TA}e^{i2q_{TA}d_0})-R_{LA,TA}R_{TA,LA}e^{2iq_{LA}d_0}e^{2iq_{TA}d_0},\qquad(3a)$$

$$S_0(\omega)=\frac{iq_{LA}}{C_{11}}\int_0^{d_0}G(\omega,z)\sin(q_{LA}z)dz.$$

In the following, if necessary, we separate in the spectral amplitudes their absolute value $S_\omega^p$ and their phase $\varphi_\omega^p$: $\tilde{S}_\omega^p=S_\omega^p\exp(i\varphi_\omega^p)$.

From Eq. (3a) we see, that $\tilde{S}_\omega^{QLA}$ and $\tilde{S}_\omega^{QTA}$ are products of two factors. The first one, $S_0$, is controlled by the properties of the laser-induced stress through $G(\omega,z)$. To calculate it exactly, it is necessary to consider self-consistently the complex processes of the nonequilibrium electron and lattice dynamics after ultra-fast laser pulse illumination. This problem is beyond the scope of this paper. Instead, we use the fact that the generated strain pulse has bipolar shape and can be approximated with good accuracy by the derivative of a Gaussian function [36]:

$$\eta_0(t-z/\upsilon_{LA})=\sqrt{e}u_{zz}^{\max}\frac{(t-z/\upsilon_{LA})}{\tau}\exp\left[-\frac{(t-z/\upsilon_{LA})^2}{2\tau^2}\right]\qquad(4a)$$

where $e$ is the base of the natural logarithm, $\tau$ is the characteristic strain pulse duration and $u_{zz}^{\max}$ is the maximal value of the only strain component $u_{zz}$ along the propagation direction. This determines the spectral form of the "seed" pulse as $S_0(\omega)=\sqrt{2\pi e}u_{zz}^{\max}\tau^2\omega\exp(-\omega^2\tau^2/2)$. The strain amplitude may be expressed through the pump excitation density $u_{zz}^{\max}=\alpha_w W$, where the coefficient $\alpha_w\sim10^{-4}$ cm$^2$/mJ [45]. For the used parameters of the optical excitation we estimate $u_{zz}^{\max}=2.8\times10^{-4}$, and the pulse duration $\tau$ remains the only fitting parameter in our consideration. In the calculations we take $\tau$=6.5 ps. Figure 1(b) shows the spatio-temporal shape of the initial pulse calculated with these parameters.

The second factor is determined by the interference of the LA and TA waves within the film and accounts for the multiple reflections and mode conversions at the interface. From Eq. (3a) we see that there are two series of resonances for which the emission into the substrate is facilitated. The frequencies of these "longitudinal" and "transverse" resonances are $f_{LA(TA)}=(1+2m)\upsilon_{LA(TA)}/(4d_0)$, where $m$ is integer. The lowest resonances are $f_{LA}\approx15$ GHz, 45 GHz and $f_{TA}\approx7.5$ GHz, 22 GHz. In our case the transmission coefficients for the processes with mode conversion are smaller than those without conversion, and the same applies for the reflection coefficients. As a result, the QLA pulse is determined mainly by the first term in Eq. (3a), and only "longitudinal" resonances are expected to be essential [see the inset in Fig. 1(c)]. For the QTA pulse, "transverse" resonances are also important, since both the first and the second terms contain reflection or transmission coefficients for the processes with mode conversion [see the inset in Fig. 1 (d)]. The spatio-temporal shapes of the strain pulses shown in Figs. 1 (c) and 1 (d) are obtained by the inverse Fourier transform of the corresponding spectra.

## Appendix B. Interaction of light with the strain pulses

The strain pulses injected into the (311) GaAs substrate perturb the initially isotropic dielectric permittivity of the substrate through the photoelastic effect and as a result modulate the intensity and polarization of the reflected probe light pulse. We assume that strain pulses propagate through the substrate and reach its free surface without change of their shape. The characteristic time of strain variation is much greater than the duration of the probe light pulse. Therefore, calculating the characteristics of the reflected light at a particular pump-probe delay, we may assume the strain (and, consequently, also the dielectric permittivity) to be static and equal to its value for this time delay. For the temporal Fourier components of the electric field inside the substrate we have

$$\frac{d^2\mathrm{E}_l}{dz'^2}=-k^2\left(1+\frac{\delta\varepsilon_{lj}}{\varepsilon_0}\right)\mathrm{E}_j, \quad l,j=x,y \qquad (1b)$$

where $\varepsilon_0=12.89$ is the dielectric permittivity of the unperturbed crystal [39] and the coordinate $z'=0$ corresponds to the open surface of the GaAs substrate. To solve this equation we apply the standard perturbation method in which the electric field $\mathrm{E}_j(z')$ is represented as the sum of the zero-strain solution of the Maxwell equations $\mathrm{E}_j^{(0)}$ and the strain-induced perturbation. $\delta\mathrm{E}_j(z')$The solution of (1b) with the standard Maxwell boundary conditions gives the resulting expression for the amplitude of the electric field modulation in the reflected light wave $\delta E_j^v$:

$$\delta \mathrm{E}_j^v = \frac{ik_0^2}{(k+k_0)} \int_{-\infty}^{0} dz' e^{-ikz'} \left[ \delta \varepsilon_{jx}(t,z') \mathrm{E}_x^{(0)}(z') + \delta \varepsilon_{jy}(t,z') \mathrm{E}_y^{(0)}(z') \right] \qquad j = x, y. \qquad (2b)$$

Since unstrained GaAs is an optically isotropic material and the normally incident light is linearly polarized, for $z' > 0$ we have $E_x^{(0)} = \cos\psi_0 [\exp(-ik_0 z') + R_{EM} \exp(ik_0 z')]$, and $E_x^{(0)} = \cos\psi_0 T_{EM} \exp(-ikz')$ for $z' < 0$, where $\psi_0$ is the angle between the incident light polarization and the $x$-axis, and $R_{EM}$ and $T_{EM}$ are the complex reflection and transmission coefficients. For $E_y^{(0)}$ we have similar expressions but with $\sin\psi_0$ instead of $\cos\psi_0$.

The perturbation of the permittivity tensor due to the strain may be expressed through the photoelastic tensor, whose standard form in crystallographic coordinates may be found, for example, in Ref. 38. For cubic crystals the photoelastic tensor has three non-zero components, $p_{11}$, $p_{12}$, and $p_{44}$. In GaAs at photon energies ~1 eV these components are $p_{11} \approx -0.165$, $p_{12} \approx -0.14$ and $p_{44} \approx -0.070$ [39,40]. In our case we must rewrite all the tensors in (311)-coordinate frame as shown in Fig. 1(a) so that for the two nonzero strain components $u_{xz}(t,z')$ and $u_{zz}(t,z')$ we come to the following expressions for the nonzero deviations of the dielectric permittivity tensor $\delta \varepsilon_{ij}(t,z')$:

$$\begin{aligned}
\delta \varepsilon_{xx}(t,z') &\approx -\varepsilon_0^2 [0.175(p_{11} - p_{12} - 2p_{44})u_{xz}(t,z') + \\
&\qquad + (0.223 p_{11} + 0.777 p_{12} - 0.446 p_{44})u_{zz}(t,z')], \\
\delta \varepsilon_{yy}(t,z') &\approx -\varepsilon_0^2 [0.386(p_{11} - p_{12} - 2p_{44})u_{xz}(t,z') + \\
&\qquad + 0.091(p_{11} + p_{12} - 2p_{44})u_{zz}(t,z')].
\end{aligned} \qquad (3b)$$

Considering the spatio-temporal dependence of the strain, we have to take into account the QLA and QTA strain pulse reflection at the free surface of GaAs. As a result the strain amplitudes in Eq. (3b) include the contributions of the incident and reflected strain pulses. Their spatial-temporal evolutions are described by Eq. (2) in the main text, but with $\eta_p(t - z/\upsilon_p)$ substituted by $\eta_p(t - t_p - z'/\upsilon_p) - \eta_p(t - t_p + z'/\upsilon_p)$. From Eq. (2b) one can obtain the expressions (3) and (4) in the main text, which are evaluated in the first-order in strain approximation. Then we come to the simplified expression (7) which is written assuming negligible light absorption ($\gamma \ll nk_0$).

41. The behavior of the Brillouin signal around $t_p$ is determined by the phase $\varphi_{as}^p$. In the case of an anti-symmetric strain pulse described by the derivative of a Gaussian function with negligible after-pulse ringing we may assume $\varphi_{as}^p = 0$ and the Brillouin signals are maximum at $t=t_p$. If the shape of the strain pulse is more complicated the behavior around $t_p$ may differ from the considered simple case as observed in Ref. 7.